\documentclass[conference]{IEEEtran}
\IEEEoverridecommandlockouts

\usepackage{cite}
\usepackage{amsmath,amssymb,amsfonts}
\usepackage{algorithmic}
\usepackage{graphicx}
\usepackage{textcomp}
\usepackage{xcolor}
\usepackage{booktabs}
\usepackage{multirow}
\usepackage{makecell}
\usepackage{diagbox}
\usepackage{tikz}
\usepackage{hyperref}
\usetikzlibrary{shapes,arrows}
\usetikzlibrary{positioning}
\usetikzlibrary {bending}
\usepackage{siunitx}
\usepackage{caption}
\usepackage{setspace}
\usepackage{tikz}

\tikzstyle{blockEntry} = [rectangle, draw, fill=red!20, 
    text width=8em, text centered, minimum height=1em]
\tikzstyle{blockConv} = [rectangle, draw, fill=blue!10, 
    text width=8em, text centered, rounded corners, minimum height=1em]
\tikzstyle{blockMaxPool} = [rectangle, draw, fill=blue!20, 
text width=8em, text centered, rounded corners, minimum height=1em]
\tikzstyle{blockUpSample} = [rectangle, draw, fill=orange!20, 
text width=8em, text centered, rounded corners, minimum height=1em]
\tikzstyle{line} = [draw, -latex']
\tikzstyle{blockEncoded} = [rectangle, draw, fill=green!20, 
    text width=8em, text centered, minimum height=1em]
\tikzstyle{blockBatch} = [rectangle, draw, fill=blue!30, 
    text width=8em, text centered, minimum height=1em]
\tikzstyle{blockTransformer} = [rectangle, draw, fill=green!50, 
    text width=8em, text centered, minimum height=1em]

\newcommand\copyrighttext{%
  \footnotesize 
  \raggedright{
  \begin{singlespace}
    \textcopyright~2025. IEEE.  Personal use of this material is permitted. Permission from IEEE must be obtained for all other uses, in any current or future media, including reprinting/republishing this material for advertising or promotional purposes, creating new collective works, for resale or redistribution to servers or lists, or reuse of any copyrighted component of this work in other works
\end{singlespace}}
    }
\newcommand\copyrightnotice{%
\begin{tikzpicture}[remember picture,overlay]
\node[anchor=south,yshift=20pt] at (current page.south) {{\parbox{\dimexpr\textwidth-\fboxsep-\fboxrule\relax}{\copyrighttext}}};
\end{tikzpicture}%
}

\def\BibTeX{{\rm B\kern-.05em{\sc i\kern-.025em b}\kern-.08em
    T\kern-.1667em\lower.7ex\hbox{E}\kern-.125emX}}
\begin{document}

\title{Planing It by Ear: Convolutional Neural Networks for Acoustic Anomaly Detection in Industrial Wood Planers}

\makeatletter
\newcommand{\linebreakand}{%
  \end{@IEEEauthorhalign}
  \hfill\mbox{}\par
  \mbox{}\hfill\begin{@IEEEauthorhalign}
}
\makeatother

\author{Anthony Deschênes$^{1,2,4,5}$, Rémi Georges$^{1,4}$, Cem Subakan$^{2,5, 7}$, Bruna Ugulino$^6$, Antoine Henry$^6$, Michael Morin$^{3,4}$ \\[0.2cm]
$^1$ Dep. of Wood and Forest Sciences, $^2$ Dep. of Computer Science and Software Engineering\\
$^3$ Dep. of Operations and Decision Systems,  
$^4$ FORAC Research Consortium\\
Université Laval, QC, Canada\\[1ex]
$^5$ Mila-Quebec AI Institute, QC, Canada, $^6$ FPInnovations, QC, Canada, $^7$ Concordia University, QC, Canada \\

}

\maketitle

\begin{abstract}
In recent years, the wood product industry has been facing a skilled labor shortage. The result is more frequent sudden failures, resulting in additional costs for these companies already operating in a very competitive market. Moreover, sawmills are challenging environments for machinery and sensors.
Given that experienced machine operators may be able to diagnose defects or malfunctions, one possible way of assisting novice operators is through acoustic monitoring.
As a step towards the automation of wood-processing equipment and decision support systems for machine operators, in this paper, we explore using a deep convolutional autoencoder for acoustic anomaly detection of wood planers on a new real-life dataset.
Specifically, our convolutional autoencoder with skip connections (Skip-CAE) and our Skip-CAE transformer outperform the DCASE autoencoder baseline, one-class SVM, isolation forest and a published convolutional autoencoder architecture, respectively obtaining an area under the ROC curve of 0.846 and 0.875 on a dataset of real-factory planer sounds. 
Moreover, we show that adding skip connections and attention mechanism under the form of a transformer encoder-decoder helps to further improve the anomaly detection capabilities.

\end{abstract}


\section{Introduction}
\copyrightnotice{}
In practice, problems and anomalies in wood-processing equipment could be detected by ear by experienced operators.
In recent years, however, the wood-product industry has been facing a skilled labor shortage. This results in an increase of abnormal behavior in wood machinery and causes additional costs to companies~\cite{buehlmann2002impact}.

One promising answer is acoustic monitoring. Acoustic monitoring is a relatively cheap and easy solution to help automatically detect these anomalies and can be seen as a tool to help novice operators to better understand their machinery. 
Acoustic anomaly detection consists of determining if a given sound segment is anomalous or not. Generally, this task is done using unsupervised learning and deep neural networks \cite{duman2020acoustic,nunesAnomalousSoundDetection2021}.
One of the challenges is that factories are noisy environments where the abnormal sounds might be damped by other noises, such as other machine operations. Moreover, some anomalies are subtle and hard to detect, even for experienced operators. Detecting such anomalies is thus a challenging task for algorithms. 

In this paper, we contribute with an industrial acoustic dataset of a functioning wood planer where anomalous events were flagged by an expert\footnote{The dataset and code are available at \href{https://github.com/AnthonyDeschenes/PlaningItByEarDataset/}{https://github.com/AnthonyDeschenes/PlaningItByEarDataset/}.}.
We present two neural network architectures for acoustic anomaly detection on this new industrial dataset---a convolutional autoencoder with skip connections (Skip-CAE), and a Skip-CAE transformer.
On our  dataset, both approaches outperform other models such as DCASE autoencoder baseline, one-class SVM, isolation forest and a convolutional autoencoder (CAE) from the literature.
To summarize, our contributions are the following:

\begin{itemize}
    \item An open industrial dataset containing real-life factory planing sounds.
    \item A new convolutional autoencoder architecture that uses skip connections to improve its learning capabilities to detect anomalies.
    \item A new convolutional autoencoder with skip connections and a transformer architecture to detect anomalies. 
    \item Comparative analysis of the two new models with existing approaches on a new industrial dataset.
\end{itemize}

\section{Background on Wood-Processing Equipment Monitoring and Acoustic Anomaly Detection}
\label{sec:back}
Wood processing equipment monitoring can be done using multiple means, such as power consumption measurement, accelerometers, displacement, temperature and acoustic emission sensors, as well as airborne sound by ultrasound sensors and microphones~\cite{mohammadpanah2019development,derbas2023multisensor}.

The potential of the acquired data is mainly focused on predicting the quality of the wood products to suggest adjustments of equipment settings~\cite{sexton2024wood,derbas2023supervised}. Regarding acoustic monitoring, most researchers in wood machining use acoustic emission sensors~\cite{nasir2019acoustic,zhuo2021overview}. However, Iskra \textit{et al.}~\cite{iskraProcessMonitoringCNC2012,iskra_comparison_2006} used PCB microphones to predict the wood surface roughness and determine the feed rate with neural networks during CNC routing. They found that other sensors such as dynamometers are inconvenient since they need to be mounted near or on the tools which can impact wood cutting negatively. Microphones are less invasive and do not affect the final product. Acoustic monitoring of wood processing equipment can include abnormal emergency event detection, often called anomaly detection, such as explosion, fire and glass breaking.

As the last machining center in the lumber production line, wood planers, complex machines equipped with at least four cutting heads with numerous knives, play an important role in adding value to lumber pieces just before the grading step. Despite the planer's degree of automation, it has no feedback system based on product quality or equipment conditions. Moreover, planer operators adjust and calibrate planer components according to their level of expertise, with little standardization from one operator to another. As a result, the planer's performance is highly dependent on the skill level of the human resources available in the plant. In addition, certain decision-making information, such as the real position of cutting tools and their wear levels, are not measured systematically or automatically. To date, there has not been enough research on monitoring of wood planing systems in the sawmilling industry~\cite{sexton2022automatic}. Hence the importance of further application of Industry 4.0 techniques, such as acoustic monitoring.

In a more general industrial setting,
Duman \textit{et al.}~\cite{duman2020acoustic} propose a deep convolutional autoencoder to detect such events/anomalies in industrial processes. They show that deep learning outperforms classical approaches such as one-class support vector machine (SVM). The one-class SVM is often used for this kind of event detection~\cite{hilal2018distributed,mnasriAnomalousSoundEvent2022,liDatadrivenSmartManufacturing2019} as well as isolation forest~\cite{antoniniSmartAudioSensors2018}. Generative adversarial neural networks are also used to detect anomalies in noisy factories~\cite{cooperAnomalyDetectionMilling2020,tagawa2021acoustic}. For acoustic anomaly detection, Nunes~\cite{nunesAnomalousSoundDetection2021} systematic review and the latest DCASE challenge~\cite{Nishida_arXiv2024_01} show that most existing approaches use mel spectograms for feature extraction from raw signals. The most common approaches used are autoencoders~\cite{oh2018residual} and convolutional autoencoders~\cite{duman2020acoustic}. 

\section{Real Planing Mill Dataset}
\label{sec:data}
Our new dataset consists of 7,562 10-second-single-channel recordings of an industrial planer (Gilbert High Speed Planer) sampled at 20 kHz for a total of approximately 21 hours of recordings with an ICP microphone model 377A21. The recordings contain real background noises coming from the planer mill and each one contains the sound of zero or more boards being planed and includes the machine start and shutdown. Some recordings have been labeled by an expert to identify anomalies. However, since the recordings come from real-life operations, some anomalies might not be labeled in the evaluation set. Table \ref{tab:dataset} presents the description of our industrial dataset with the number of data for each type of board according to the North American Lumber Grading Standards  (2 inches by 3 inches, 2 by 4 and 2 by 6). The data are split in a training set and a evaluation set. The training set contains recordings with no anomalies that were recorded shortly after knives jointing operations or head change. It thus contains data representing normal planer operating sounds. The evaluation set data span two days of operation at a real planing mill, each day is started with new cutting heads. Evaluation data contains 105 anomalies---four broken boards passing through the planer, 29 boards being stuck in the planer, and 72 uneven or thick boards going through the planer. All of them result in different abnormal sounds. 

\begin{table}[h]
\caption{Dataset description with the number of recordings per board type and the total number of anomalies per set}
\resizebox{\columnwidth}{!}{%
\begin{tabular}{lllll}
\toprule
             & \thead{\# 2$\times$3} & \thead{\# 2$\times$4} & \thead{\# 2$\times$6} & \thead{\# Anomalies} \\
\midrule
Training set & 90     & 1897   & 2340   & 0            \\
Evaluation set     & 0      & 0      & 3235   & 105            \\
\bottomrule
\end{tabular}%
}
\label{tab:dataset}
\end{table}

Each data has been pre-processed into mel spectograms with a frame size of 50 ms with a hop length of 25 ms and 80 mel bins. As a result, each data contains 32,080 values.

\section{Building the Anomaly Detection Models}
\label{sec:models}
For our neural networks, we built autoencoders, and convolutional autocencoders.
The \emph{autoencoder}~\cite{Harada_EUSIPCO2023_01} is a deep feedforward neural network described as the baseline in the DCASE challenge of 2024~\cite{Nishida_arXiv2024_01} (note that the DCASE MobileNetV2 baseline is not included since our dataset contains only one section). It has three parts: the encoder, the bottleneck and the decoder. The encoder has four fully connected layers of 128 neurons. The bottleneck has one layer of eight neurons and the decoder has four fully connected layers of 128 neurons. There is batch normalization between each layer. All activation functions are ReLU. We changed its input and output 
for mel spectograms leading to 32,080 neurons.

The \emph{convolutional autoencoder} (CAE) model was introduced by Duman \textit{et al.}~\cite{duman2020acoustic}. Its encoder has 10 convolutional 3 by 3 layers and four max pooling 2 by 2 layers. Its decoder has 12 convolutional 3 by 3 layers and four upsampling 2 by 2 layers. All activation functions are ReLU. We changed the entry size to fit the size of the mel spectograms of our dataset and the number of channels after each convolution has been reduced by a factor of four to obtain a more compact latent representation. The upsamplings are done using bilinear upsampling.

We also use a \emph{convolutional autoencoder with skip connections} (Skip-CAE) inspired by the CAE of Duman \textit{et al.}~\cite{duman2020acoustic}.
Figure~\ref{fig:proposed_conv_architecture} presents its architecture. We used leaky ReLU---instead of ReLU---to improve the convergence and prevent vanishing gradient. Moreover, we added batch normalization before each max pooling and upsampling layers. When the number of channels is not the same between two layers of a skip connection (for example the entry that has one channel and the first batch normalization that has four channels), the skip connection is only done on the minimum number of channels (first channel for the example).

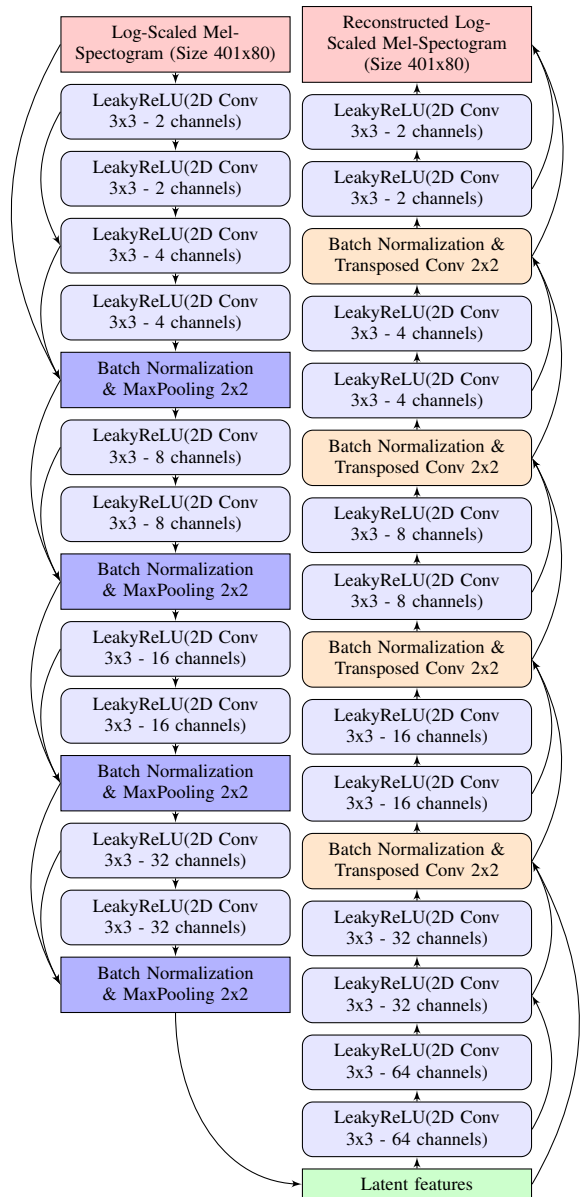
\begin{figure}
\center

\begin{tikzpicture}[node distance = 1cm, auto]
\tikzstyle{every node}=[font=\scriptsize]
    \node [blockEntry] (init) {Log-Scaled Mel-Spectogram (Size 401x80)};
    \node [blockConv, below=0.15cm of init] (conv1_1) {LeakyReLU(2D Conv 3x3 - 2 channels)};
    \node [blockConv, below=0.15cm of conv1_1] (conv1_2) {LeakyReLU(2D Conv 3x3 - 2 channels)};
    \node [blockConv, below=0.15cm of conv1_2] (conv2_1) {LeakyReLU(2D Conv 3x3 - 4 channels)};
    \node [blockConv, below=0.15cm of conv2_1] (conv2_2) {LeakyReLU(2D Conv 3x3 - 4 channels)};
    \node [blockBatch, below=0.15cm of conv2_2] (batch1) {Batch Normalization \& MaxPooling 2x2};
    \node [blockConv, below=0.15cm of batch1] (conv3_1) {LeakyReLU(2D Conv 3x3 - 8 channels)};
    \node [blockConv, below=0.15cm of conv3_1] (conv3_2) {LeakyReLU(2D Conv 3x3 - 8 channels)};
    \node [blockBatch, below=0.15cm of conv3_2] (batch2) {Batch Normalization \& MaxPooling 2x2};
    \node [blockConv, below=0.15cm of batch2] (conv4_1) {LeakyReLU(2D Conv 3x3 - 16 channels)};
    \node [blockConv, below=0.15cm of conv4_1] (conv4_2) {LeakyReLU(2D Conv 3x3 - 16 channels)};
    \node [blockBatch, below=0.15cm of conv4_2] (batch3) {Batch Normalization \& MaxPooling 2x2};
    \node [blockConv, below=0.15cm of batch3] (conv5_1) {LeakyReLU(2D Conv 3x3 - 32 channels)};
    \node [blockConv, below=0.15cm of conv5_1] (conv5_2) {LeakyReLU(2D Conv 3x3 - 32 channels)};
    \node [blockBatch, below=0.15cm of conv5_2] (batch4) {Batch Normalization \& MaxPooling 2x2};

    \node [blockEntry, right= 0.15cm of init] (end) {Reconstructed Log-Scaled Mel-Spectogram (Size 401x80)};
    \node [blockConv, below=0.15cm of end] (conv11_2) {LeakyReLU(2D Conv 3x3 - 2 channels)};
    \node [blockConv, below=0.15cm of conv11_2] (conv11_1) {LeakyReLU(2D Conv 3x3 - 2 channels)};
    \node [blockUpSample, below=0.15cm of conv11_1] (up4) {Batch Normalization \& Transposed Conv 2x2};
    \node [blockConv, below=0.15cm of up4] (conv10_2) {LeakyReLU(2D Conv 3x3 - 4 channels)};
    \node [blockConv, below=0.15cm of conv10_2] (conv10_1) {LeakyReLU(2D Conv 3x3 - 4 channels)};
    \node [blockUpSample, below=0.15cm of conv10_1] (up3) {Batch Normalization \& Transposed Conv 2x2};
    \node [blockConv, below=0.15cm of up3] (conv9_2) {LeakyReLU(2D Conv 3x3 - 8 channels)};
    \node [blockConv, below=0.15cm of conv9_2] (conv9_1) {LeakyReLU(2D Conv 3x3 - 8 channels)};
    \node [blockUpSample, below=0.15cm of conv9_1] (up2) {Batch Normalization \& Transposed Conv 2x2};
    \node [blockConv, below=0.15cm of up2] (conv8_2) {LeakyReLU(2D Conv 3x3 - 16 channels)};
    \node [blockConv, below=0.15cm of conv8_2] (conv8_1) {LeakyReLU(2D Conv 3x3 - 16 channels)};
    \node [blockUpSample, below=0.15cm of conv8_1] (up1) {Batch Normalization \& Transposed Conv 2x2};
    \node [blockConv, below=0.15cm of up1] (conv7_2) {LeakyReLU(2D Conv 3x3 - 32 channels)};
    \node [blockConv, below=0.15cm of conv7_2] (conv7_1) {LeakyReLU(2D Conv 3x3 - 32 channels)};
    \node [blockConv, below=0.15cm of conv7_1] (conv6_2) {LeakyReLU(2D Conv 3x3 - 64 channels)};
    \node [blockConv, below=0.15cm of conv6_2] (conv6_1) {LeakyReLU(2D Conv 3x3 - 64 channels)};

    \node [blockEncoded, below=0.15cm of conv6_1] (encoded) {Latent features};
    
    \path [line] (init) -- (conv1_1);
    \path [line] (conv1_1) -- (conv1_2);
    \path [line] (conv1_2) -- (conv2_1);
    \path [line] (conv2_1) -- (conv2_2);
    \path [line] (conv2_2) -- (batch1);
    \path [line] (init.west) to [out=240,in=120] (batch1.west); 
    \path [line] (conv1_1.west) to [out=240,in=120] (conv2_1.west);
    \path [line] (conv2_1.west) to [out=240,in=120] (batch1.west);
    \path [line] (batch1) -- (conv3_1);
    \path [line] (conv3_1) -- (conv3_2);
    \path [line] (conv3_2) -- (batch2);
    \path [line] (batch1.west) to [out=240,in=120] (batch2.west);
    \path [line] (conv3_1.west) to [out=240,in=120] (batch2.west);
    \path [line] (batch2) -- (conv4_1);
    \path [line] (conv4_1) -- (conv4_2);
    \path [line] (conv4_2) -- (batch3);
    \path [line] (batch2.west) to [out=240,in=120] (batch3.west);
    \path [line] (conv4_1.west) to [out=240,in=120] (batch3.west);
    \path [line] (batch3) -- (conv5_1);
    \path [line] (conv5_1) -- (conv5_2);
    \path [line] (conv5_2) -- (batch4);
    \path [line] (batch3.west) to [out=240,in=120] (batch4.west);
    \path [line] (conv5_1.west) to [out=240,in=120] (batch4.west);
    \path [line] (batch4.south) to [out=-90,in=180] (encoded.west);
    \path [line] (encoded) -- (conv6_1);
    \path [line] (conv6_1) -- (conv6_2);
    \path [line] (conv6_1.east) to [out=60,in=-60] (conv7_1.east);
    \path [line] (conv7_1.east) to [out=60,in=-60] (up1.east);
    \path [line] (encoded.east) to [out=60,in=-60] (up1.east);
    \path [line] (conv6_2) -- (conv7_1);
    \path [line] (conv7_1) -- (conv7_2);
    \path [line] (conv7_2) -- (up1);
    \path [line] (up1) -- (conv8_1);
    \path [line] (conv8_1) -- (conv8_2);
    \path [line] (up1.east) to [out=60,in=-60] (up2.east);
    \path [line] (conv8_1.east) to [out=60,in=-60] (up2.east);
    \path [line] (conv8_2) -- (up2);
    \path [line] (up2) -- (conv9_1);
    \path [line] (conv9_1) -- (conv9_2);
    \path [line] (up2.east) to [out=60,in=-60] (up3.east);
    \path [line] (conv9_1.east) to [out=60,in=-60] (up3.east);
    \path [line] (conv9_2) -- (up3);
    \path [line] (up3) -- (conv10_1);
    \path [line] (conv10_1) -- (conv10_2);
    \path [line] (conv10_2) -- (up4);
    \path [line] (up3.east) to [out=60,in=-60] (up4.east);
    \path [line] (conv10_1.east) to [out=60,in=-60] (up4.east);
    \path [line] (up4) -- (conv11_1);
    \path [line] (conv11_1) -- (conv11_2);
    \path [line] (up4.east) to [out=60,in=-60] (end.east);
    \path [line] (conv11_1.east) to [out=60,in=-60] (end.east);
    \path [line] (conv11_2) -- (end);
\end{tikzpicture}
\caption{Architecture of the CAE with skip connections}
\label{fig:proposed_conv_architecture}
\vspace*{-1.7\baselineskip}
\end{figure}  

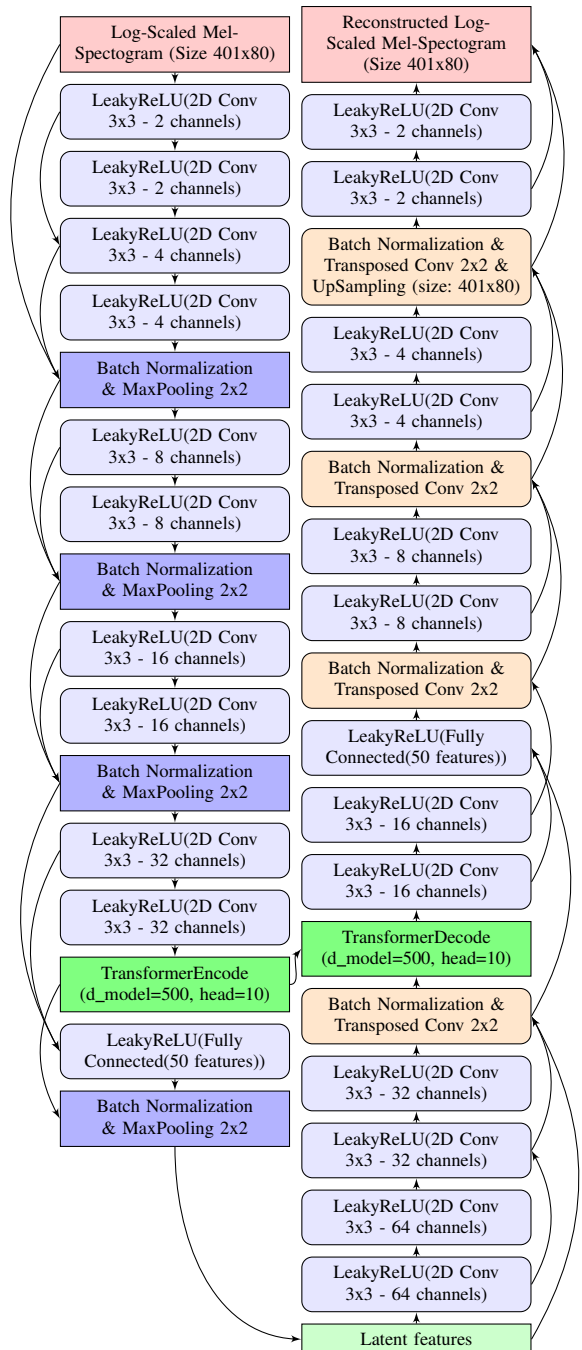
\begin{figure}

\begin{tikzpicture}[node distance = 1cm, auto]
\tikzstyle{every node}=[font=\scriptsize]
    \node [blockEntry] (init) {Log-Scaled Mel-Spectogram (Size 401x80)};
    \node [blockConv, below=0.15cm of init] (conv1_1) {LeakyReLU(2D Conv 3x3 - 2 channels)};
    \node [blockConv, below=0.15cm of conv1_1] (conv1_2) {LeakyReLU(2D Conv 3x3 - 2 channels)};
    \node [blockConv, below=0.15cm of conv1_2] (conv2_1) {LeakyReLU(2D Conv 3x3 - 4 channels)};
    \node [blockConv, below=0.15cm of conv2_1] (conv2_2) {LeakyReLU(2D Conv 3x3 - 4 channels)};
    \node [blockBatch, below=0.15cm of conv2_2] (batch1) {Batch Normalization \& MaxPooling 2x2};
    \node [blockConv, below=0.15cm of batch1] (conv3_1) {LeakyReLU(2D Conv 3x3 - 8 channels)};
    \node [blockConv, below=0.15cm of conv3_1] (conv3_2) {LeakyReLU(2D Conv 3x3 - 8 channels)};
    \node [blockBatch, below=0.15cm of conv3_2] (batch2) {Batch Normalization \& MaxPooling 2x2};
    \node [blockConv, below=0.15cm of batch2] (conv4_1) {LeakyReLU(2D Conv 3x3 - 16 channels)};
    \node [blockConv, below=0.15cm of conv4_1] (conv4_2) {LeakyReLU(2D Conv 3x3 - 16 channels)};
    \node [blockBatch, below=0.15cm of conv4_2] (batch3) {Batch Normalization \& MaxPooling 2x2};
    \node [blockConv, below=0.15cm of batch3] (conv5_1) {LeakyReLU(2D Conv 3x3 - 32 channels)};
    \node [blockConv, below=0.15cm of conv5_1] (conv5_2) {LeakyReLU(2D Conv 3x3 - 32 channels)};
    \node [blockTransformer, below=0.15cm of conv5_2] (transf_encode) {TransformerEncode (d\_model=500, head=10)};
    \node [blockConv, below=0.15cm of transf_encode] (lin_transformer_encode) {LeakyReLU(Fully Connected(50 features))};
    \node [blockBatch, below=0.15cm of lin_transformer_encode] (batch4) {Batch Normalization \& MaxPooling 2x2};

    \node [blockEntry, right= 0.15cm of init] (end) {Reconstructed Log-Scaled Mel-Spectogram (Size 401x80)};
    \node [blockConv, below=0.15cm of end] (conv11_2) {LeakyReLU(2D Conv 3x3 - 2 channels)};
    \node [blockConv, below=0.15cm of conv11_2] (conv11_1) {LeakyReLU(2D Conv 3x3 - 2 channels)};
    \node [blockUpSample, below=0.15cm of conv11_1] (up4) {Batch Normalization \& Transposed Conv 2x2 \& UpSampling (size: 401x80)};
    \node [blockConv, below=0.15cm of up4] (conv10_2) {LeakyReLU(2D Conv 3x3 - 4 channels)};
    \node [blockConv, below=0.15cm of conv10_2] (conv10_1) {LeakyReLU(2D Conv 3x3 - 4 channels)};
    \node [blockUpSample, below=0.15cm of conv10_1] (up3) {Batch Normalization \& Transposed Conv 2x2};
    \node [blockConv, below=0.15cm of up3] (conv9_2) {LeakyReLU(2D Conv 3x3 - 8 channels)};
    \node [blockConv, below=0.15cm of conv9_2] (conv9_1) {LeakyReLU(2D Conv 3x3 - 8 channels)};
    \node [blockUpSample, below=0.15cm of conv9_1] (up2) {Batch Normalization \& Transposed Conv 2x2};
    \node [blockConv, below=0.15cm of up2] (lin_transf_decode) {LeakyReLU(Fully Connected(50 features))};
    \node [blockConv, below=0.15cm of lin_transf_decode] (conv8_2) {LeakyReLU(2D Conv 3x3 - 16 channels)};
    \node [blockConv, below=0.15cm of conv8_2] (conv8_1) {LeakyReLU(2D Conv 3x3 - 16 channels)};
    \node [blockTransformer, below=0.15cm of conv8_1] (transf_decode) {TransformerDecode (d\_model=500, head=10)};
    \node [blockUpSample, below=0.15cm of transf_decode] (up1) {Batch Normalization \& Transposed Conv 2x2};
    \node [blockConv, below=0.15cm of up1] (conv7_2) {LeakyReLU(2D Conv 3x3 - 32 channels)};
    \node [blockConv, below=0.15cm of conv7_2] (conv7_1) {LeakyReLU(2D Conv 3x3 - 32 channels)};
    \node [blockConv, below=0.15cm of conv7_1] (conv6_2) {LeakyReLU(2D Conv 3x3 - 64 channels)};
    \node [blockConv, below=0.15cm of conv6_2] (conv6_1) {LeakyReLU(2D Conv 3x3 - 64 channels)};

    \node [blockEncoded, below=0.15cm of conv6_1] (encoded) {Latent features};
    
    \path [line] (init) -- (conv1_1);
    \path [line] (conv1_1) -- (conv1_2);
    \path [line] (conv1_2) -- (conv2_1);
    \path [line] (conv2_1) -- (conv2_2);
    \path [line] (conv2_2) -- (batch1);
    \path [line] (init.west) to [out=240,in=120] (batch1.west); 
    \path [line] (conv1_1.west) to [out=240,in=120] (conv2_1.west);
    \path [line] (conv2_1.west) to [out=240,in=120] (batch1.west);
    \path [line] (batch1) -- (conv3_1);
    \path [line] (conv3_1) -- (conv3_2);
    \path [line] (conv3_2) -- (batch2);
    \path [line] (batch1.west) to [out=240,in=120] (batch2.west);
    \path [line] (conv3_1.west) to [out=240,in=120] (batch2.west);
    \path [line] (batch2) -- (conv4_1);
    \path [line] (conv4_1) -- (conv4_2);
    \path [line] (conv4_2) -- (batch3);
    \path [line] (batch2.west) to [out=240,in=120] (batch3.west);
    \path [line] (conv4_1.west) to [out=240,in=120] (batch3.west);
    \path [line] (batch3) -- (conv5_1);
    \path [line] (conv5_1) -- (conv5_2);
    \path [line] (conv5_2) -- (transf_encode);
    \path [line] (transf_encode.west) to [out=240,in=120] (batch4.west);
    \path [line] (batch3.west) to [out=240,in=120] (lin_transformer_encode.west);
    \path [line] (conv5_1.west) to [out=240,in=120] (lin_transformer_encode.west);
    \path [line] (batch4.south) to [out=-90,in=180] (encoded.west);
    \path [line] (lin_transformer_encode) -- (batch4);
    
    \path [line] (encoded) -- (conv6_1);
    \path [line] (conv6_1) -- (conv6_2);
    \path [line] (conv6_1.east) to [out=60,in=-60] (conv7_1.east);
    \path [line] (conv7_1.east) to [out=60,in=-60] (up1.east);
    \path [line] (encoded.east) to [out=60,in=-60] (up1.east);
    \path [line] (conv6_2) -- (conv7_1);
    \path [line] (conv7_1) -- (conv7_2);
    \path [line] (conv7_2) -- (up1);
    \path [line] (transf_decode) -- (conv8_1);
    \path [line] (up1) -- (transf_decode);
    \path [line] (conv8_1) -- (conv8_2);
    \path [line] (up1.east) to [out=60,in=-60] (lin_transf_decode.east);
    \path [line] (conv8_1.east) to [out=60,in=-60] (lin_transf_decode.east);
    \path [line] (conv8_2.east) to [out=60,in=-60] (up2.east);
    \path [line] (lin_transf_decode) -- (up2);
    \path [line] (up2) -- (conv9_1);
    \path [line] (conv9_1) -- (conv9_2);
    \path [line] (up2.east) to [out=60,in=-60] (up3.east);
    \path [line] (conv9_1.east) to [out=60,in=-60] (up3.east);
    \path [line] (conv9_2) -- (up3);
    \path [line] (up3) -- (conv10_1);
    \path [line] (conv10_1) -- (conv10_2);
    \path [line] (conv10_2) -- (up4);
    \path [line] (up3.east) to [out=60,in=-60] (up4.east);
    \path [line] (conv10_1.east) to [out=60,in=-60] (up4.east);
    \path [line] (up4) -- (conv11_1);
    \path [line] (conv11_1) -- (conv11_2);
    \path [line] (up4.east) to [out=60,in=-60] (end.east);
    \path [line] (conv11_1.east) to [out=60,in=-60] (end.east);
    \path [line] (conv11_2) -- (end);
    \path [line] (transf_encode.east) to [out=0,in=180] (transf_decode.west);
\end{tikzpicture}
\caption{Architecture of the CAE with skip connections and transformer}
\label{fig:proposed_conv_transform_architecture}
\vspace*{-1.7\baselineskip}
\end{figure}

We also use a \emph{convolutional autoencoder with skip connections and transformer} (Skip-CAE-Transformer).
Figure \ref{fig:proposed_conv_transform_architecture} presents its architecture. Compared to the CAE with skip connections of Figure~\ref{fig:proposed_conv_architecture}, we added a TransformerEncode block and a TransformerDecode block to the model. Each of these block has 10 heads and a single layer. These transformers are the pytorch implementation of Vaswani \textit{et al.}~\cite{vaswaniAttentionAllYou}. The connection between the TransformerEncode and the TransformerDecode represents the memory of the transformer. Moreover, the skip connections before the last batch normalization and the second transposed convolution pass through a fully connected layer to let the network decide if the values of these connections are interesting or not. This model can also be seen as a means to reduce the dimensionality of the input before using a transformer architecture.

All neural networks are trained for 500 epochs using  Adam with decoupled weight decay (AdamW) optimizer~\cite{loshchilovDecoupledWeightDecay2019a}. 
The same hyperparamters are used for training:
a batch size of 32, an initial learning rate of \num{1e-3}, a cosine scheduler~\cite{loshchilov2017sgdr} with a minimum learning rate of \num{1e-5}, a maximum learning rate of \num{1e-3} and five warm-up restart steps. An early stopping is done if the validation loss has not improved for 30 epochs and the best validation loss model is restored at the end of the training. We used the mean squared error (MSE) loss, defined as the average of the square of the differences between the data values for the mel spectogram and its reconstruction. We use 10\% of the training set as the validation set.

We also built models using one-class SVM and isolation forest.
\emph{One-class SVM}, called OneClassSVM in scikit-learn~\cite{scikit-learn}, is a commonly used anomaly detection algorithm. 
Apart from using a RBF kernel, all parameters are set to the default scikit-learn values. 
\emph{Isolation forest} is a common anomaly detection algorithm based on random forest. 
We use the scikit-learn implementation with 100 estimators and default values for the other parameters~\cite{scikit-learn}.

\section{Experiments: Evaluating the Models}
\label{sec:experiments}
The main objective of the experiments is to compare and evaluate the proposed models. 
Similar to the DCASE challenge, we use two metrics: the area under the receiver operating characteristic curve (AUC ROC), and the partial AUC (pAUC) with $p = 0.1$. 
The reason for using the pAUC is that an anomaly detector that has many false positives cannot be trusted. 
All models are trained on an Intel Core i7-8750H CPU @ 2.20~GHz, 6 cores, 16~GB of RAM and an NVIDIA GeForce GTX 1050 GPU with 4 GB of memory.

\subsection{Results for all anomalies}
Figure~\ref{fig:roc_curves} presents the ROC curves of the different models. The legend shows the AUC and pAUC of each model. Isolation forest is the worst performing model with an AUC (pAUC) of \num{0.518} (\num{0.474}). The DCASE baseline performs poorly with an AUC of \num{0.520} (\num{0.477}). One-class SVM also performs poorly with an AUC of \num{0.683} (\num{0.516}). Compared with the CAE of Duman \textit{et al.}~\cite{duman2020acoustic}, the Skip-CAE performs significantly better with an AUC of \num{0.846} (\num{0.787}) compared to \num{0.798} (\num{0.720}). The Skip-CAE-Transformer performs the best with an AUC of \num{0.875} (\num{0.785}). Skip-CAE and Skip-CAE-Transformer are the only two models that can identify anomalies without false positives with a true positive rate of 20\%. Figure~\ref{fig:reconstruction_abnormal_data} shows a mel spectogram corresponding to a broken board anomaly. It also shows the reconstructed mel spectograms of the different neural network models and their reconstruction errors. Our Skip-CAE and Skip-CAE-Transformer show a deeper understanding of the data compared to the baseline networks. For example, the sound of each planed board is clearly distinguishable in the reconstructed mel spectograms of our models while it is not for the two other models.

\begin{figure}[htbp]
\centerline{\includegraphics[width=\columnwidth]{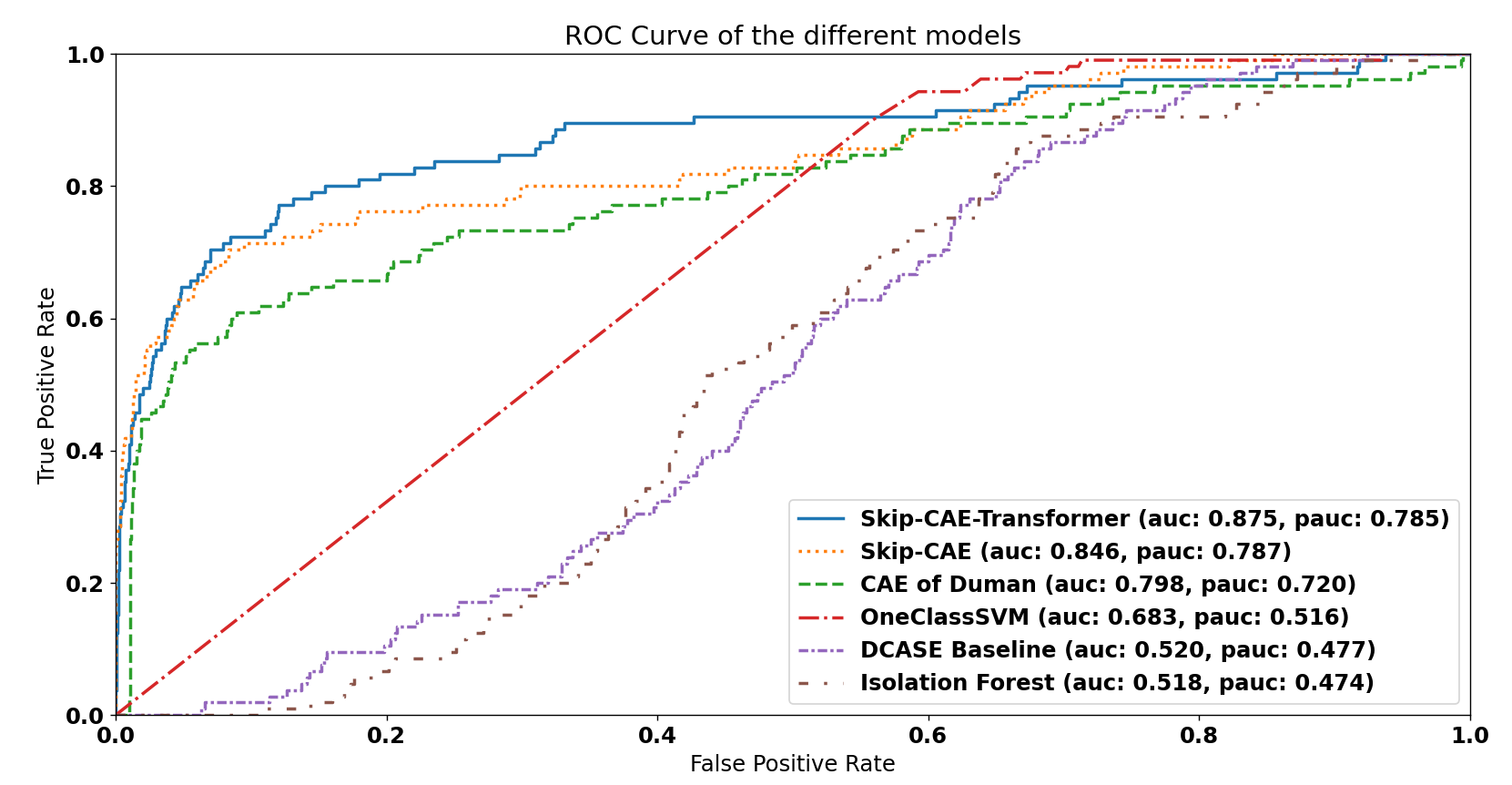}}
\caption{ROC curves, AUC and pAUC of the different models on the evaluation set}
\label{fig:roc_curves}
\end{figure}

\begin{figure}[htbp]
\centerline{\includegraphics[width=\columnwidth]{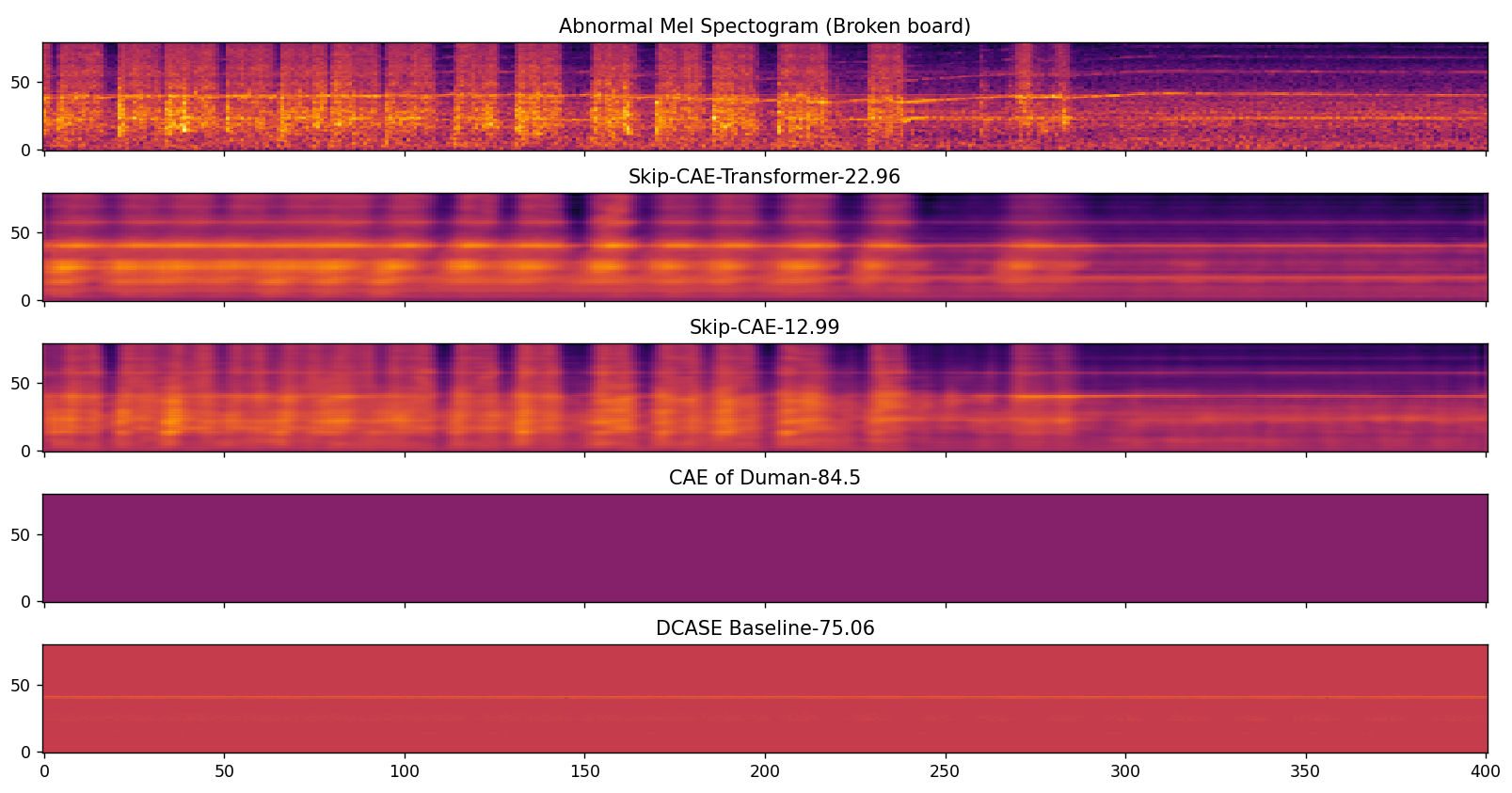}}
\caption{Mel spectogram of an anomaly (broken board) and the reconstructed mel spectograms of the neural network models with their reconstruction errors}
\label{fig:reconstruction_abnormal_data}
\end{figure}

\subsection{Results per anomaly type}
Table \ref{tab:score_per_anomaly_type} presents the AUC and pAUC of the different models for each type of anomaly. The best result for each metric is in bold. The CAE with skip connections and transformer (Skip-CAE-Transformer) outperforms all other models in AUC or pAUC for all of the anomaly types. 

\begin{table}[h!]
\caption{AUC and pAUC score of each model for each anomaly type}
\resizebox{\columnwidth}{!}{%
\begin{tabular}{@{}lcccccc@{}}
\toprule
\diagbox[outerleftsep=0.2mm]{\textbf{Model}}{\textbf{Anomaly Type}} & \multicolumn{2}{l}{\textbf{Broken Board}} & \multicolumn{2}{l}{\textbf{Board Stuck}} & \multicolumn{2}{l}{\textbf{Uneven or Thick Wood}} \\ 
                 & AUC            & pAUC           & AUC   & pAUC  & AUC   & pAUC  \\
                 \midrule
Skip-CAE-Transformer  & 0.743               & \textbf{0.677}               & \textbf{0.778}      & \textbf{0.744}     & \textbf{0.921}          & 0.807          \\
Skip-CAE         & \textbf{0.777} & 0.618          & 0.723 & 0.729 & 0.900 & \textbf{0.820} \\
CAE of Duman \textit{et al.}~\cite{duman2020acoustic}    & 0.757          & 0.644 & 0.639 & 0.676 & 0.864 & 0.742 \\
OneClassSVM      & 0.558          & 0.509          & 0.707 & 0.519 & 0.681 & 0.515 \\
DCASE Baseline   & 0.441          & 0.474          & 0.620 & 0.480 & 0.484 & 0.476 \\
Isolation Forest & 0.437          & 0.474          & 0.613 & 0.474 & 0.485 & 0.474 \\ \bottomrule
\end{tabular}%
}
\label{tab:score_per_anomaly_type}
\end{table}

\section{Conclusion}
\label{sec:conclusion}
We developed two deep neural network models inspired by a CAE introduced by Duman \textit{et al.}~\cite{duman2020acoustic} to detect anomalies in audio signals from an industrial wood planer. 
The first model, Skip-CAE, is a CAE with skip connections to improve the convergence.
The second, Skip-CAE-Transformer, has additional transformer encode and decode layers.
For the experiments, we built a new public dataset of real-life-recorded wood planer operations containing 105 anomalies corresponding to stuck boards, broken boards and uneven or thick boards. 
We compared, on this dataset, the proposed neural networks to baselines such as one-class SVM, isolation forest, the CAE of Duman \textit{et al.}~\cite{duman2020acoustic}, and a traditional autoencoder used in the DCASE 2024 challenge. 
Our best model, the Skip-CAE-Transformer, outperforms the other models obtaining an AUC of \num{0.875}. 
Grouping results by anomaly type reveals that the Skip-CAE-Transformer outperforms other models to detect stuck boards and uneven or thick wood. Our models thus demonstrate that it is feasible to detect abnormal events on industrial planers. Furthermore, acoustic anomaly detection is close to how operators function. As such, our models have the potential to gain the trust of expert operators who can understand their outcomes, but it can also help new operators understand the reason for the alarms/anomalies detected.

\addtolength{\textheight}{-8.0cm}   

\bibliographystyle{IEEEtran}
\bibliography{biblio}

\end{document}